\documentclass[pre,twocolumn,a4paper,superscriptaddress,groupedaddress,notitlepage]{revtex4-2}
\usepackage{amssymb}
\usepackage{amsmath,bm}
\usepackage{graphicx,color}
\usepackage{multirow}
\usepackage[normalem]{ulem}
\usepackage{dcolumn}

\newcommand{\B}[1]{{\bm{#1}}}
\newcommand{\R}[1]{{\textrm{#1}}}

\begin{document}

\title{Shear-induced mixing of granular materials featuring broad granule size distributions}
\author{Joyjit Chattoraj}
\author{Nguyen Hoang Huy}
\author{Saurabh Aggarwal}
\author{Mohamed Salahuddin Habibullah}
\author{Farzam Farbiz}
\address{Institute of High Performance Computing, Agency for Science Technology and Research, \#16-16 Connexis, 1 Fusionopolis Way, Singapore 138632}

\date{\today}

\begin{abstract}
Granular flows during a shear-induced mixing process are studied using Discrete Element Methods. The aim is to understand the underlying elementary mechanisms of transition from unmixed to mixed phases for a granular material featuring a broad distribution of particles, which we investigate systematically by varying the strain rate and system size. Here the strain rate varies over four orders of magnitude and the system size varies from ten thousand to more than a million granules. A strain rate-dependent transition from quasistatic to purely inertial flow is observed. 
At the macroscopic scale, the contact stresses drop due to the formation of shear-induced instabilities that serves as an onset of  granular flows and initiates mixing between the granules.
The stress-drop displays a profound system size dependence. At the granular scale, mixing dynamics are correlated with the formation of shear bands, which result in significantly different timescales of mixing, especially for those regions that are close to the system walls and the bulk. 
Overall, our results reveal that although the transient dynamics display a generic behavior these have a significant finite-size effect. In contrast, macroscopic behaviors at steady states have negligible system size dependence.   
\end{abstract}

\maketitle

\section{Introduction}
The mixing of granular particles has long been recognized as a complex process which affects the performance of many products in many industries including pharmaceutical, food and cosmetic. The outcomes of a mixing process, or more generically referred to as the flow properties, markedly depend on the particle size distribution, density, roughness, moisture and cohesion~\cite{Campbell2011,VlachosChang2011,MellmannHF2013,OpalinskiCH2016,Vivacquaetal2019}, as well as externally supplied energy~\cite{Argentina2002,Claric2008}. It is also well-known reality that particles with similar properties tend to segregate themselves from others~\cite{Rhodes2008}. For granular particles, they also exhibit percolation, whereby small particles under the influence of gravity ends up staying at the bottom of a test-bed, or they percolate under the influence of shear. Experiments as well as computer simulations have explored shear-induced percolation extensively.
Many different industrially relevant geometries are examined such as ball mills, blenders, hoppers~\cite{Cleary2009, Sudahetal2005, Anandetal2008}. The results obtained from such complex geometries are crucial, however, these could be geometry specific. The complexity in geometries poses a problem to rationalize the pure response of particle properties from the geometric effects.

Over the years, especially in discrete element methods (DEM), minimal geometries such as simple shear or plane shear have been extensively employed either via Lees-Edwards boundary conditions or via simulation walls to explore the elementary mechanisms of flow behaviors~\cite{Campbell1989,LoisLC2005,CruzEPRC2005,XuOHern2006,Guoetal2013,MandalKhakhar2018}. 
These geometries have also been utilized to understand the shear-induced transition dynamics of mixing in granular materials.   
Lu and Hsiau~\cite{LuHsiau2008} imposed shear through rough parallel walls and studied the mixing process of a binary assembly starting from an unmixed phase. They concluded that the mixing is governed by diffusion mechanisms. 
Aarons et al~\cite{AaronsBH2013} adopted a similar sheared geometry with a binary mixture of significantly high diameter ratio of 7:1, and investigated the role of interparticle cohesive force on mixing. They found that the homogeneity of the binary mixture is correlated with the cohesive forces between small particles, even though locally they did not find any correlation between solid volume fraction and homogeneity.
 In addition, Obreg\'on et al~\cite{Obregonetal2010a, Obregonetal2010b} in their periodic shear experiments found that a faster mixing is achieved with increasing particle size as the large diameter reduces the frictional contact area. Similarly, they found that particles near the moving walls mix better than the rest of the system. 

One major drawback in using DEM simulations is in its inability to experimentally match the relevant scales of mixing due to limited computational power. The standard procedures to overcome such computational shortcomings involve either enlarging the particle-sizes by a constant factor, or shortening the simulation dimensions~\cite{KholaWassgren2016}. The majority of the DEM studies on mixing focused on the binary assemblies of similar size-order. In real industrial products, however, this is not the case as the sizes of the granular particles are often distributed over a few orders of magnitude~\cite{Ambrose2016}. In these materials, such adhoc dimensional adjustments need to be implemented carefully as the wide distributions of sizes introduce different length-scales and time-scales during the mixing processes. Thus, an in-depth understanding on how to set an optimum simulation scale to study the mixing  of industrially relevant granular products has a significant commercial impact, but it has thus far received little research attention.  


In this study, we investigate systematically the finite-size effect of mixing in a cohesion-less granular assembly featuring a broad distribution of particle sizes.  We begin from an unmixed phase and shear the system using a pair of parallel flat walls, forcing the granules to be mixed. We then study the transition dynamics from unmixed to mixed phases for several strain rates in the absence of gravity. We show that the shear-induced transition dynamics display qualitatively similar behavior, however, quantitatively the transient behavior depends on the system size. Upon further examination of the steady state regimes of the granular flows, we show that the macroscopic variables are insensitive to the finite size effect. Results obtained from this study display a significant influence of shear bands on mixing which is crucial in developing a continuum model for granular mixing.  

The rest of the paper is organised in following sections. In the next section, we describe the granular model and details of the DEM simulations. In the following section, we discuss the mixing dynamics, effects of finite-size and effects of strain rates on mixing and steady state properties. Finally, we summarize the key findings in the summary section.             

\section{Model and methodology}
\subsection{Model specification}
Inspired from industrially relevant powders~\cite{Ambrose2016}, we prepare a polydisperse granular assembly containing seven different types of granules which are of significantly different sizes and masses, and present with unique weight proportions (Table~\ref{Table:powder}). We select a cubic box with a volume fraction of 0.6, whereby volume fraction is defined as the ratio of the total volume occupied by the granules to the box volume. We prepare three different system sizes where the total number of particles are 13147, 131441, 1314378, respectively. 
\begin{table}
\caption{Characteristics of granular particles.}
\label{Table:powder}
\begin{ruledtabular}
  \begin{tabular}{cccc}
 type & diameter & mass & weight(\%) \\ 
 \hline
  1 & 1.0   & 1.0  &  0.3\\
  2 & 1.23  & 2.504 & 1.5\\
  3 & 1.944 & 5.161 & 4.7\\
  4 & 2.212 & 4.12 & 55.4 \\ 
  5 & 3.998 & 41.85 & 27\\
  6 & 4.269 & 48.16 & 10.8\\
  7 & 21.39 & 13157 & 0.2\\
  
\end{tabular}    
\end{ruledtabular}
\end{table}

\subsection{DEM modeling}
In our modelling, we use the open source software LAMMPS to perform discrete element simulations for the shear-induced mixing studies~\cite{Plimpton1995}. Here the particles are assumed to be soft spheres. Two particles exert a contact force only when they overlap. The overlapped distance $\delta_{ij}$ is measured as the difference between the sum of radii $R_i+R_j$ and the center-to-center distance $r_{ij}$, i.e., $\delta_{ij}=(R_i + R_j - r_{ij})$. The pair vector ${\B r}_{ij}=\B r_i-\B r_j$ is the vector difference between the two position vectors $\B r_i$ and $\B r_j$. 

The contact force $\B F_{ij}$ is comprised of a normal force $\B F_{\R n}$ term and a tangential force $\B F_{\R t}$ term. The normal force, $\B F_{\R n} = \B F_{\R n\R e} +\B F_{\R n\R d}$, is further comprised of an elastic force $\B F_{\R n\R e}$ and a damping force $\B F_{\R n \R d}$. Both the forces are acting along the normal direction of the pair $\hat{n}_{ij} = \B r_{ij}/r_{ij}$. We express the elastic force using the Hertzian model  
\begin{equation}
    {\B F}_{\R n\R e} = k_{n}\sqrt{\frac{R_iR_j}{R_i+R_j}} \delta_{ij}^{3/2}\hat{n}_{ij} ,
    \label{Eq:hertz}
\end{equation}
where $k_n$ is the stiffness coefficient of the interparticle bond, having the dimension of pressure with the value set at $17\times10^6$. The damping force is proportional to the relative velocity between particle $i$ and particle $j$  
\begin{equation}
    {\B F}_{\R n\R d} = -\eta_{n}({\B v}_i - {\B v}_j)\cdot \hat{n}_{ij}\hat{n}_{ij}, 
    \label{Eq:ndamp}
\end{equation}
where $\eta_{n}$ is the damping coefficient which follows the relation
\begin{equation}
    \eta_{n} = \alpha \sqrt{ \frac{m_i m_j}{m_i + m_j} k_n \sqrt{\frac{R_iR_j}{R_i+R_j} \delta_{ij}}}, 
    \label{Eq:ndampc}
\end{equation}
proposed by Tsuji et al~\cite{TsujiTI1992}. Here $m_i$ and $m_j$ are the masses of particle $i$ and $j$, and $\alpha$ is a constant, whose value depends on the coefficient of restitution $e$, set to 0.5. 

Similar to the normal force, the tangential force, or more precisely the friction force, $\B F_{\R t}$ is also comprised of two forces, the Mindlin force $\B F_{\R t\R m}$~\cite{Mindlin1949} and a damping force $\B F_{\R t\R d}$, both acting along the tangential direction with respect to the pair vector $\B r_{ij}$. The tangential force 
\begin{equation}
    \B F_{\R t} = -\min\left(\mu |\B F_{\R n}|, |\B F_{\R t\R m} + \B F_{\R t\R d}|\right) \hat{t}_{ij},
    \label{Eq:tforce}
\end{equation}
reaches the Coulomb limit when the sum of Mindlin and damping force overcomes a fraction of the normal force. The limit is controlled by the coefficient of friction $\mu$, a scalar quantity whose value is set to 0.5 in this study. $\hat{t}_{ij}$ is the unit vector along the tangential direction. The Mindlin force 
\begin{equation}
    \B F_{\R t\R m} = -k_t \sqrt{\frac{R_iR_j}{R_i+R_j}}(\delta_{ij})^{1/2}\xi_{ij}\hat{t}_{ij}, 
    \label{Eq:mindlin}
\end{equation}
depends on both the overlapped distance $\delta_{ij}$ and the tangential distance $\xi_{ij}$ which further depends on the contact history. The tangential displacement $\B\xi_{ij}=\int_{t_0}^{t_p} \B v_{\R t_{ij}}(t') \mathrm{d}t'$ is the integration of the relative tangential velocity $\B v_{\R t_{ij}}$ over time ranging from the initial time of the contact formation $t_0$ to the present time $t_p$. The history is erased and $\xi_{ij}$ is set to zero once the contact breaks. The tangential stiffness parameter, $k_t$, is set to $6.5\times10^6$. The tangential damping force    
\begin{equation}
    \B F_{\R t\R d} = -0.9\eta_{n} \B v_{\R t_{ij}},
    \label{Eq:tdamp}
\end{equation}
is proportional to $\B v_{\R t_{ij}}$, which includes translation velocities in the form of relative velocity  $\B v_{ij}=\B v_i -\B v_j$ along $\B t_{ij}$, and the rotation velocities $\omega_i$ and $\omega_j$, and it is defined as 
\begin{equation}
    \B v_{\R t_{ij}} =  \B v_{ij} - (\B v_{ij}\cdot\hat{n}_{ij})\hat{n}_{ij}- (R_i\B\omega_i + R_j\B\omega_j)\times \hat{n}_{ij}.
    \label{Eq:velt}
\end{equation}

\begin{figure}
    \begin{center}
        \includegraphics[width=0.4\textwidth]{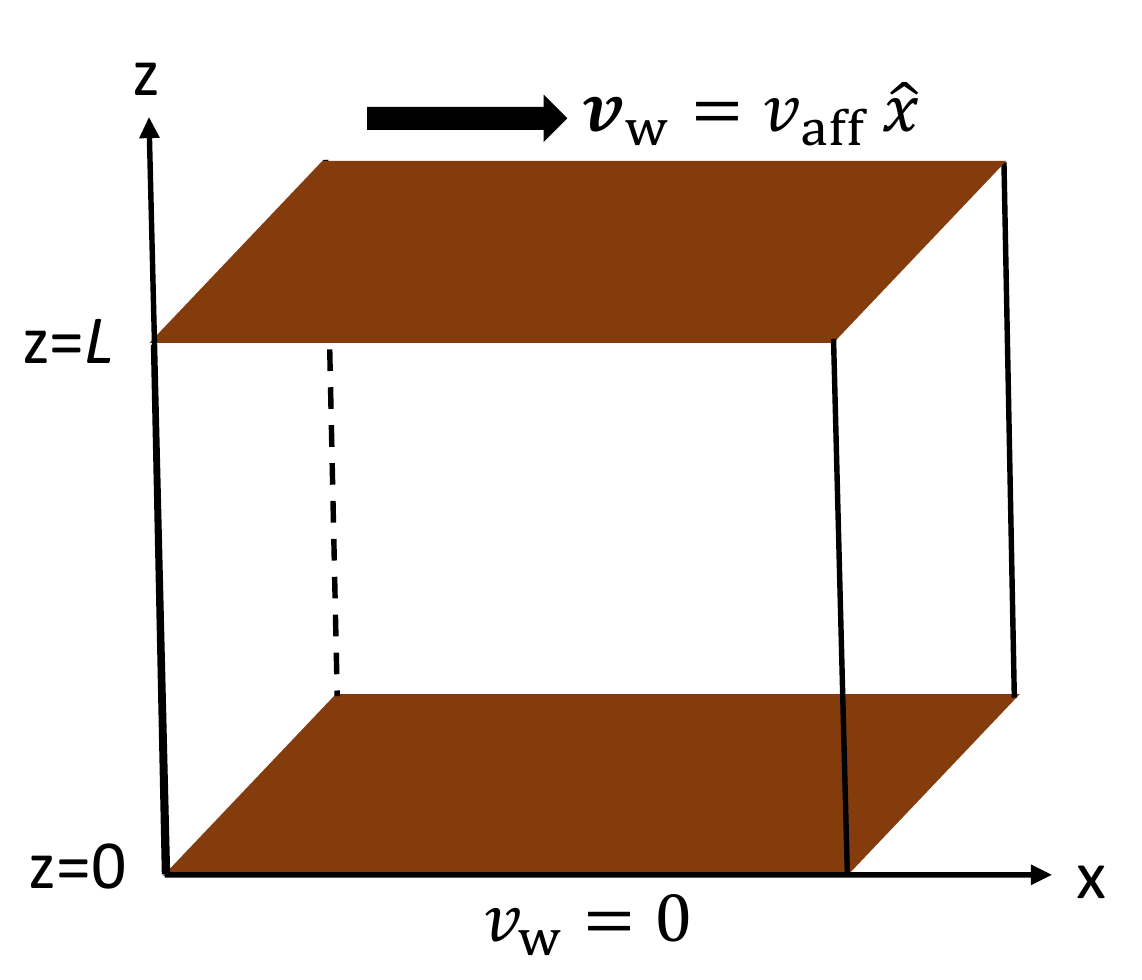}
        \caption{A schematic of a cubic simulation box of length $L$. The top and bottom planes are covered by two smooth flat walls. The top wall moves with a constant velocity and the bottom wall is static, that generates a shear flow inside the simulation box.  
        }
        \label{Fig:cartoon}
    \end{center}
\end{figure} 
From the schematic of the cubic simulation box in Fig.~\ref{Fig:cartoon}, it can be deduced that the granular particles also experience contact forces from the two parallel flat walls, implemented at the bottom ($\B r_{w}=\B 0$) and at the top ($\B r_w=L\hat{z}$) of the cubic simulation box (length $L$), to perform simple shear operations. The other two dimensions, x and y, are not bounded, implying that the particles can move back and forth based on the periodic boundary conditions. The top wall moves with a constant velocity $\B v_{w}=v_{\R{aff}}\hat{x}$ along the horizontal direction, while the bottom wall is static, i.e. $\B v_w=\B 0$.  
The particle which comes into contact with the top wall will experience a shear force due to $\B v_{\R {aff}}$. Thus, the velocity facilitates generating a simple shear profile in the simulation box across the z-axis that can be described by an imposed strain rate 
\begin{equation}
\dot{\gamma}=v_{\R {aff}}/L. 
\end{equation}
It is, however, a known fact that the shear profile will not be uniform. 
Rather than being a constant $\dot\gamma$ profile, the system will have a z-dependent strain rate profile~\cite{AaronsSundaresan2006}.
It is possible to achieve a uniform simple shear profile using the Lees-Edwards boundary conditions, but this behavior is not observable experimentally~\cite{LiMc2006}. 
Note that with increasing system size $L$, we increase $v_{\R{aff}}$ accordingly such that the imposed $\dot{\gamma}$ remains constant.

When a particle $i$ comes into contact with a wall $w$, the wall exerts: (i) a normal force
\begin{equation}
\B F_{\R n\R w}=\left[ k_n\delta_{iw} - 50m_i(\B v_i-\B v_w) \cdot \hat{n}_{iw} \right](R_i\delta_{iw})^{1/2}\hat{n}_{iw}, 
\label{Eq:fnw}
\end{equation}
and (ii) a tangential force 
\begin{equation}
\B F_{\R t\R w}=-\min(\mu|\B F_{\R n\R w}|,|\B F_{\R t\R m\R w}|)\hat{t}_{iw}
\label{Eq:ftw}
\end{equation}
on the particle, where $\B F_{\R t\R m\R w}=-k_t(R_i\delta_{iw})^{1/2}\xi_{iw}\hat{t}_{iw}$ is the Mindlin force, $\delta_{iw}=R_i-r_{iw}$ is the overlap distance, $r_{iw}$ is the magnitude of the vector $\B r_{iw} = z_i \hat{z} - \B r_w$, $z_i$ is the z-position of the particle. Here, all the parameters carry the same meanings and values as defined in equations~(\ref{Eq:hertz},\ref{Eq:tforce},\ref{Eq:mindlin}).   

Thus, the net force $\B F_i$ acting on a particle $i$ is the sum of all contact forces due to pair contacts $\B F_{ij}$ and due to walls $\B F_{iw}$. In addition, a damping force, which is proportional to particle-velocity $-0.01\B v_{i}$, is added to $\B F_i$. The damping term restricts a particle to free flow, drains the kinetic energy out of the system, and prevents the occurrence of any oscillatory instability~\cite{ChattorajGPCP2019a, ChattorajGPCP2019b}.  
Similarly, the particle $i$ experiences a net torque $\B T_{i}$ which is the accumulation of torques induced by tangential pair forces~(\ref{Eq:tforce}) $(R_i-\delta_{ij}/2)(\B F_{\R t}\times \hat{n}_{ij})$, and torques induced by tangential wall forces~(\ref{Eq:ftw}) $R_i(\B F_{\R {tw}}\times \hat{n}_{iw})$. The net force and torque acting on a particle generate translation and rotational motions, which are numerically estimated over time using the standard velocity-Verlet algorithm. In this study, the time step $\delta t = 10^{-5}$ is selected for numerical integration to ensure stable numerical computations. 

In our DEM simulations, the cut-off distance is set equal to the largest particle size in building the neighbor list, it results in long computational times for the neighbors and long communication times between the parallel processors. However, the maximum number of neighbors per particle on average is about 5, which is small because of our implementation of a repulsive interaction. We have used 120 parallel CPUs (Intel Xeon CPU E5-2690 $2.6$GHz) to simulate our largest system consisting of more than one-million particles, which take approximately $0.011$s to compute one smallest time step, addressing an important consideration in computational efficiency and effectiveness.   

The fundamental mass ($m$) and length ($d$) units in our simulations are defined by the mass and diameter of the smallest particle (see Table~\ref{Table:powder}) and the time unit ($t_{\R{col}}$) is defined by 
\begin{equation}
t_{\R{col}}= \sqrt{\frac{m}{d k_n}}.
\label{Eq:tcol}
\end{equation}
\begin{figure*}
    \begin{center}
        \includegraphics[width=0.95\textwidth]{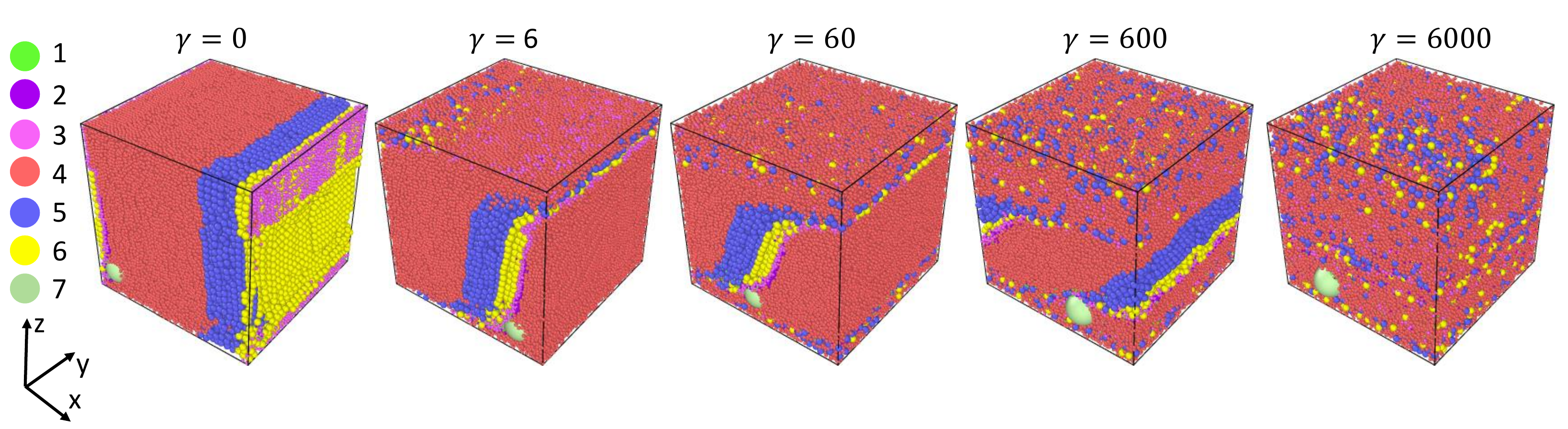}\\
        \includegraphics[width=0.95\textwidth]{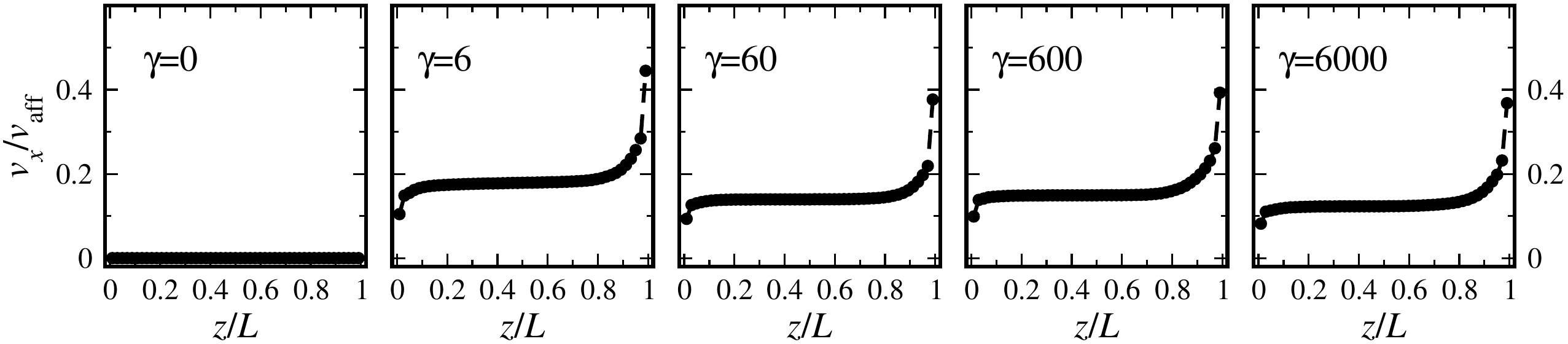}
        \caption{Granular flows at different stages of shear $\gamma=\dot\gamma \times \R{time}$ (top). Here the simulation box has length $L=114.94$, contains 131441 particles, and it is sheared at a constant strain rate $\dot{\gamma} t_\R{col}=2.4\times 10^{-5}$. 
        $v_x$, the average velocity of particles along $x$, as a function of $z$ is plotted for the above configurations (bottom).    
        \label{Fig:configs}
        }
    \end{center}
\end{figure*}

\subsection{Preparation protocol}
We prepare an initial configuration where seven granule types are stacked side by side, representing an unmixed granular phase. 
During the stacking process we keep the box elongated along the x-axis, whilst the other two dimensions are kept constant at value $L$ and start stacking each type of granule one after another. 
We checked that the choice of stacking-orders of granular types has mere influence on the results reported in this article.
Once the stacking is completed, we gently shrink the x-dimension to length $L$. Next, we bring the configuration at mechanical equilibrium, i.e., the net force and torque acting on each particle become negligible. 
An unmixed configuration prepared using the above protocol can be seen in the top left panel of Fig.~\ref{Fig:configs}. 
Once the initial configuration is prepared, we switch on the velocity $v_{\R{aff}}$ of the top wall of the simulation box and start our shear-induced mixing study.    

\subsection{Local variables}
\label{SubSec:local}
Experimentally, the mixing status is evaluated by using a statistical index, which determines the homogeneity of the particle distribution. Normally a few sub-domains from the entire system are chosen for statistical sampling to conduct the homogeneity test. The dimension of the sub-domain is then carefully selected such that it is much larger than the typical size of particles but smaller than the experimental system~\cite{AsachiNH2018}.      

In this study, we divide the simulation box into equal cubes of dimension $27$, which is $26\%$ larger than the largest particle. The length is large enough to accommodate all types of granules with right proportions. The mixing status at each cube is then quantified by examining the relative density fluctuations of the granules with respect to the global density. We define a dimensionless parameter $\delta\Theta$ as follows  
\begin{equation}
\delta\Theta = \frac{1}{N_{\R{type}}}\sum_{s=1}^{N_{\R{type}}}\left|\frac{\rho_{\R{loc};s}}{\rho_{\R{loc}}}-\frac{\rho_s}{\rho}\right|,
    \label{Eq:doh}
\end{equation}
where $\rho$ and $\rho_s$ are the global densities for all granules and only for granules of type-$s$, respectively. Similarly, $\rho_{\R{loc}}$ and $\rho_{\R{loc};s}$ are the local densities for all granules and only for granules of type-$s$ at the test cube, respectively. $N_{\R{type}}$ is the total types of granules present in the system. 
The ideal mixing proportions of the granules at the test cube correspond to values of $\delta\Theta$ equal to zero, whereas the non-zero values of $\delta\Theta$ indicate deviations from the ideal.
In addition, we calculate the local strain rate for each test cube
\begin{equation}
\dot\gamma_{\R{loc}} = \frac{1}{N_{\R{loc}}}\sum_{i=1}^{N_{\R{loc}}} \frac{v_{x;i}}{z_i},
    \label{Eq:gdoteff}
\end{equation}
where $N_{\R{loc}}$ is the number of granules in the cube, $z_i$ and $v_{x;i}$ are the z-position and x-velocity of granule $i$.
Similarly, we calculate the coordination number for each test cube
\begin{equation}
n_{\R{c}} = \frac{1}{N_{\R{loc}}}\sum_{i=1}^{N_{\R{loc}}} n_{{\R c};i},
    \label{Eq:coordination}
\end{equation}
where $n_{{\R c};i}$ is the number of granules in contact with granule $i$.

Furthermore, in order to study the response of the three local variables against shear, we compute the average $\delta\Theta$, $\dot\gamma_{\R{loc}}$ and $n_{\R{c}}$ over all of the cubes lying on the same z-plane, denoted as $\langle\delta\Theta\rangle$, $\langle\dot\gamma_{\R{loc}}\rangle$ and $\langle n_{\R{c}}\rangle$, respectively. 

\section{Results and discussions}
\subsection{Dynamics of mixing}
From the initial unmixed condition, particles move as layers along the horizontal direction with shear strain $\gamma=\dot\gamma \times \R{time}$ (see Fig.~\ref{Fig:configs}): It can be clearly observed that the particles close to the top wall move fast as the wall moves with a constant velocity $v_{\R{aff}}$, whilst those particles at the bottom are the slowest, which is unsurprising as the bottom wall is static, so the wall-friction opposes the sheared motion. A layer formation across the z-axis is also apparent when looking at the blue particles in the top panel, $\gamma=6$. 
Eventually, those particles close to the top and bottom walls start to mix (panels $\gamma=60,600$). The particles in the middle are mixed later after applying a large amount of shear (panel $\gamma=6000$).   

Such layerwise motion can be quantified through the study of velocity profile as a function of $z$. Here, we divide the z-dimension of the box into several layers with equal thicknesses. 
Next, we compute the average x-velocity of the particles $v_x$ for each layer. 
$v_x$ rescaled by $v_{\R{aff}}$ as a function of $z/L$ is displayed in the bottom row of Fig.~\ref{Fig:configs}.
The mechanical equilibrium state of the initial configuration results in zero velocity (panel $\gamma=0$). Once the shearing process starts, a non-zero velocity profile is established. Theoretically, a straight line velocity profile with slope one is expected for simple shear. 
Instead, we find a non-linear velocity profile signifying the formation of shear bands, typically observed in amorphous and jammed systems~\cite{SchallvanHecke2010, Divouxetal2016}. 
Here, the non-linear velocity profile can be characterized into three regimes, namely, the top, bottom, and middle or bulk regimes. The bulk regime appears nearly flat with varying $z/L$ as the particles move with a constant velocity. The magnitude of the velocity at the bulk is smaller than $v_\R{aff}$, and it further drops with increasing $\gamma$. In comparison to the bulk, the other two regimes display sharp non-zero slopes. The bottom regime appears always shorter than the top regime. 
A similar type of shear band formation in granular assemblies was previously reported by Shojaaee et al~\cite{ShojaaeeRCW2012}.

\begin{figure}
    \begin{center}
        \includegraphics[width=0.45\textwidth]{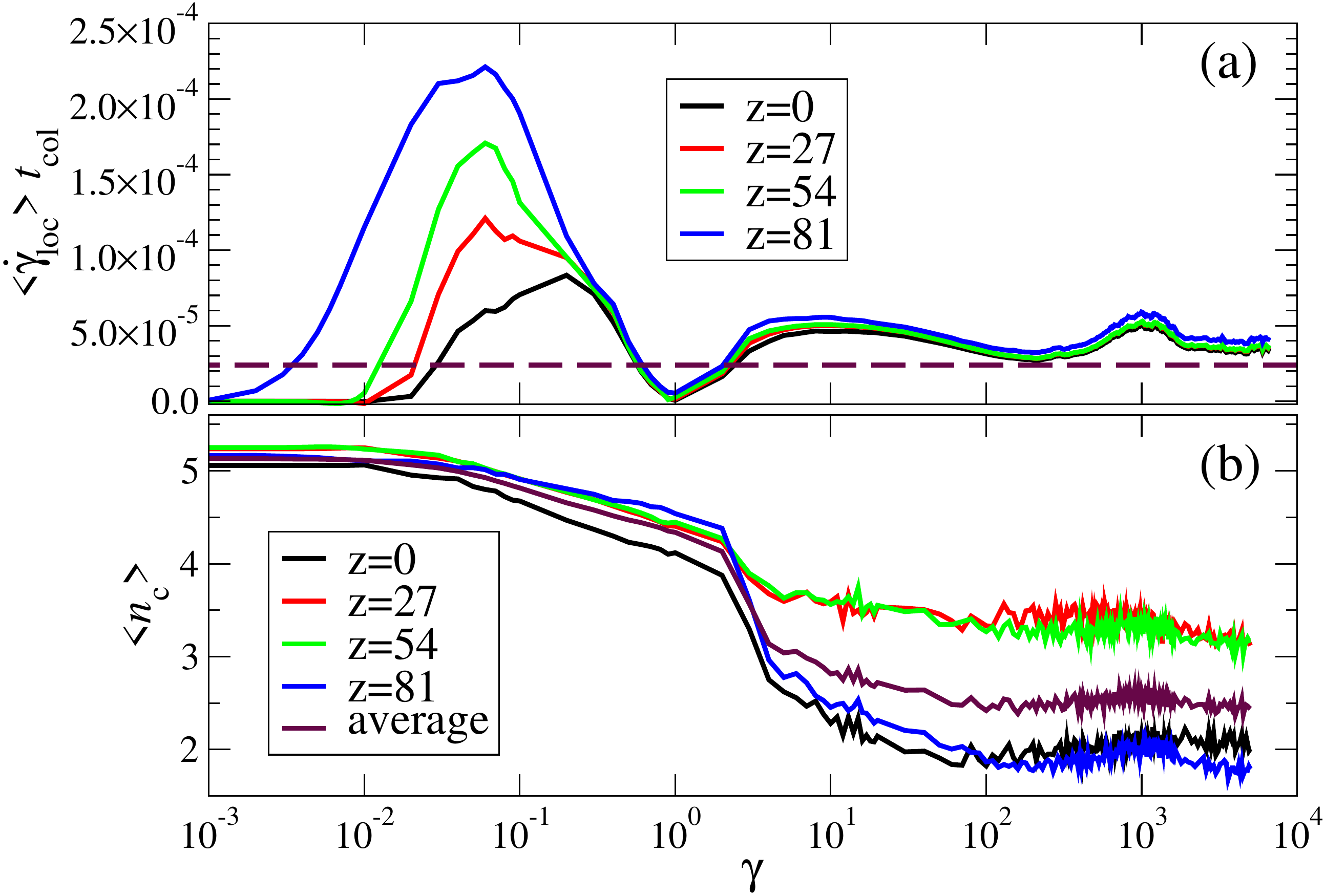}
        \caption{ (a) The average local strain rate $\langle\dot\gamma_{\R{loc}}\rangle t_{\R{col}}$ of test cubes lying on the same z-plane is shown as a function over $\gamma$ for $N=131441$. The dashed line indicates the imposed strain rate $\dot\gamma t_{\R{col}}=2.4\times 10^{-5}$. 
        (b) Similarly, the average coordination number $\langle n_{\R{c}}\rangle$ is plotted over $\gamma$. 
        }
        \label{Fig:gdoteff}
    \end{center}
\end{figure}

The existence of shear bands restricts the system in having a uniform strain rate profile locally. In understanding the local strain rate behavior over strain, we study the average local strain rate $\langle\dot\gamma_{\R{loc}}\rangle$, as defined in section~\ref{SubSec:local}. 
In Fig.~\ref{Fig:gdoteff}(a), we discover an oscillatory profile for $\langle\dot\gamma_{\R{loc}}\rangle$, saturating to a plateau at large strain intervals. It can also be observed that the oscillations have a systematic z-dependence, with the oscillation-amplitude increasing with increasing z. Even in the plateau regime, we find that $\langle\dot\gamma_{\R{loc}}\rangle$ increases with z, and
all of the plateau values are higher than the imposed strain rate $\dot\gamma$.

On further examination, we also find that the average coordination number $\langle n_{\R{c}}\rangle$ dramatically drops over strain, and it is sensitive with shear bands (Fig.~\ref{Fig:gdoteff}(b)). It is also evident that at large strain intervals, $\langle n_{\R{c}}\rangle$ displays two different regimes, i.e., one close to the walls, and the other at the bulk.       


The above results suggest that the granular system evolves from unmixed to mixed states under shear through the formation of shear bands.
Recently, Vasisht and Del Gado numerically showed a strong correlation between shear banding and macroscopic properties, for example pressure in soft systems such as emulsions~\cite{VasishtDelGado2020}. The authors established that the evolution of pressure over strain, which is marked by a plastic yielding followed by a steady state response, is associated with the emergence, disappearance and characteristic shape changes of shear bands.
We further investigate if there exists any correlation between the pressure and mixing dynamics.

We follow the standard definition of pressure $P$~\cite{ThompsonPM2009}
\begin{equation}
    P = \frac{N}{L^3}T + \frac{1}{3}(\sigma_{\R{xx}} + \sigma_{\R{yy}} + \sigma_{\R{zz}}),
    \label{Eq:P}
\end{equation}
where the first term on the right hand side is the kinetic term, $N$ is the total number of particles and $T$ is the kinetic energy from both the translation and rotation
\begin{equation}
\begin{split}
    T = \frac{2}{6N}\Bigg[\frac{1}{2} \sum_{i=1}^Nm_i\{(v_{ix}-v_x(z_i&))^2 + v_{iy}^2 + v_{iz}^2\}\\
      & + \frac{1}{4}\sum_{i=1}^Nm_iR_i^2\omega_i^2 \Bigg],
\end{split}
\label{Eq:T}
\end{equation}
where we remove the bias velocity $v_{x}(z_i)$ from the x-velocity $v_{ix}$ of each particle. 
$v_{x}(z_i)$ is determined by averaging the x-velocities of particles within the same z-plane as particle $i$.

The second term on the right hand side of~(\ref{Eq:P}) is the average contribution of the three normal stress components of the virial stress tensor $\sigma_{\alpha\beta}$, which follows the definition
\begin{equation}
    \sigma_{\alpha\beta} = \frac{1}{L^3} \left[ \sum_{i=1; i\neq j}^{N'} r^{\alpha}_{ij} F^{\beta}_{ij} + \sum_{i=1}^{N'} r^{\alpha}_{iw} F^{\beta}_{iw} \right],
    \label{Eq:sigma}
\end{equation}
where $N'$ is the sum of the particles in the simulation box ($N$) and the image particles~\cite{AllenTildesley2017} within close proximity to the simulation box.
$r^{\alpha}_{ij}$ is the $\alpha$-component of the pair vector either between two particles, or between a particle and an image particle, or between two image particles. 
Similarly, $F^{\beta}_{ij}$ is the $\beta$-component of the pair force between $i$ and $j$. The stress due to the walls is included in the second term on the right hand side, and the suffix $w$ represents the two walls.

\begin{figure}
    \begin{center}
        \includegraphics[width=0.45\textwidth]{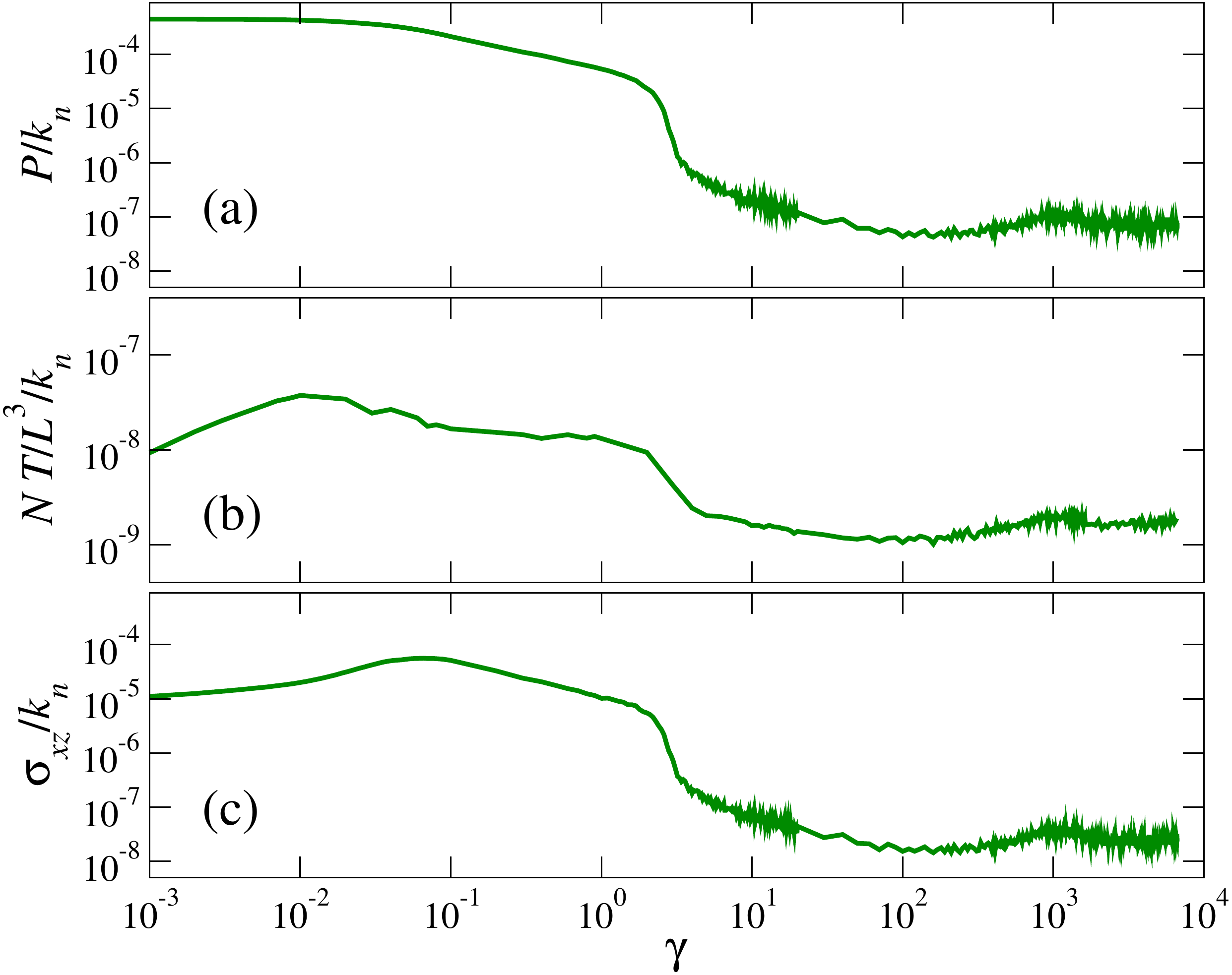}
        \caption{Pressure $P$ (a), kinetic component of $P$ (b), and shear stress $\sigma_{xz}$ (c) rescaled by $k_n$ and plotted over $\gamma$ for $N=131441$ and $\dot{\gamma} t_\R{col}=2.4\times 10^{-5}$.   
        \label{Fig:P}
        }
    \end{center}
\end{figure}
\begin{figure}
    \begin{center}
        \includegraphics[width=0.45\textwidth]{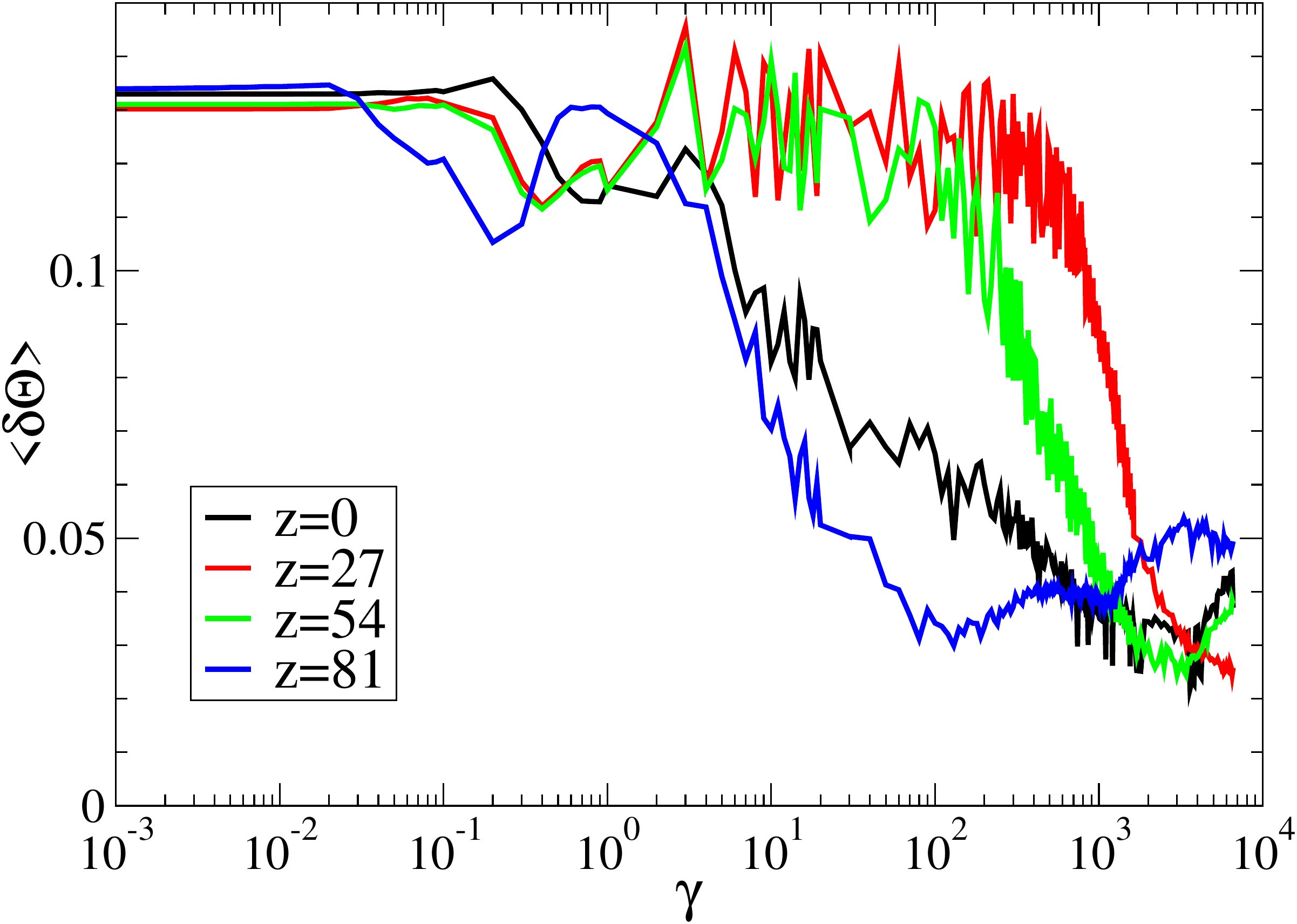}
        \caption{$\langle\delta\Theta\rangle$, the density fluctuation parameter averaged over cubes locating at the same $z$-plane, versus $\gamma$ for $\dot\gamma t_\R{col}=2.4\times 10^{-5}$ and $N=131441$.
        }
        \label{Fig:doh}
    \end{center}
\end{figure}

The results in Fig.~\ref{Fig:P} show that the initial value of $P$ is high. The $P$-value decreases weakly until a sharp drop takes place, then $P$ displays a non-monotonic behavior, and finally, it reaches a steady state at around $\gamma \sim 3\times 10^3$ as shown in Fig.~\ref{Fig:P}(a). We further confirm the steady state by conducting another simulation starting from a configuration where the particles are placed randomly. The random system also displays the same plateau value for pressure under shear. A similar sharp drop around the same $\gamma$ can be seen in both kinetic pressure (panel (b)) and shear stress $\sigma_{xz}$ (panel (c)), and both the quantities reach plateaus also around the same $\gamma$ as $P$. The kinetic contribution on pressure is negligible as compared to the contribution from the contact forces, whilst the other two components of shear stress $\sigma_{xy}$ and $\sigma_{yz}$, as expected, fluctuates around zero with vanishing magnitude.   

The sharp drop in $P$ represents a static to dynamic transition of the whole system. The transition initiates granular flow, and thus, particles start to mix, which can be seen in Fig.~\ref{Fig:doh}.  
We study the average density fluctuation parameter $\langle{\delta\Theta}\rangle$, defined in section~\ref{SubSec:local}, for several $z$-planes. $\langle\delta\Theta\rangle$ at the bottom $z=0$ and near the top walls $z=81$ show a jump occurring around the same $\gamma$ when the $P$ value  drops. 
These low values of $\langle{\delta\Theta}\rangle$ indicate that the granular particles in these two planes reach the mixed phases stage. 
$\langle\delta\Theta\rangle$ at the bulk takes longer shear deformation for mixing to occur. It is also evident that the strain value at which $\langle\delta\Theta\rangle$ drops increases with the distance from the top wall. 
These results from our studies show similar mixing behaviors characteristic as observed in the periodic shear experiments~\cite{Obregonetal2010a} where the granular particles close to the two walls were also found to be mixed fast compared to the bulk particles. Here, we found that the fast mixing occurs due to the presence of the two shear bands which create slip events. Slipping randomizes particle motion along x, y and z directions.

\subsection{Effects of finite-size}
\begin{figure}
    \begin{center}
        \includegraphics[width=0.4\textwidth]{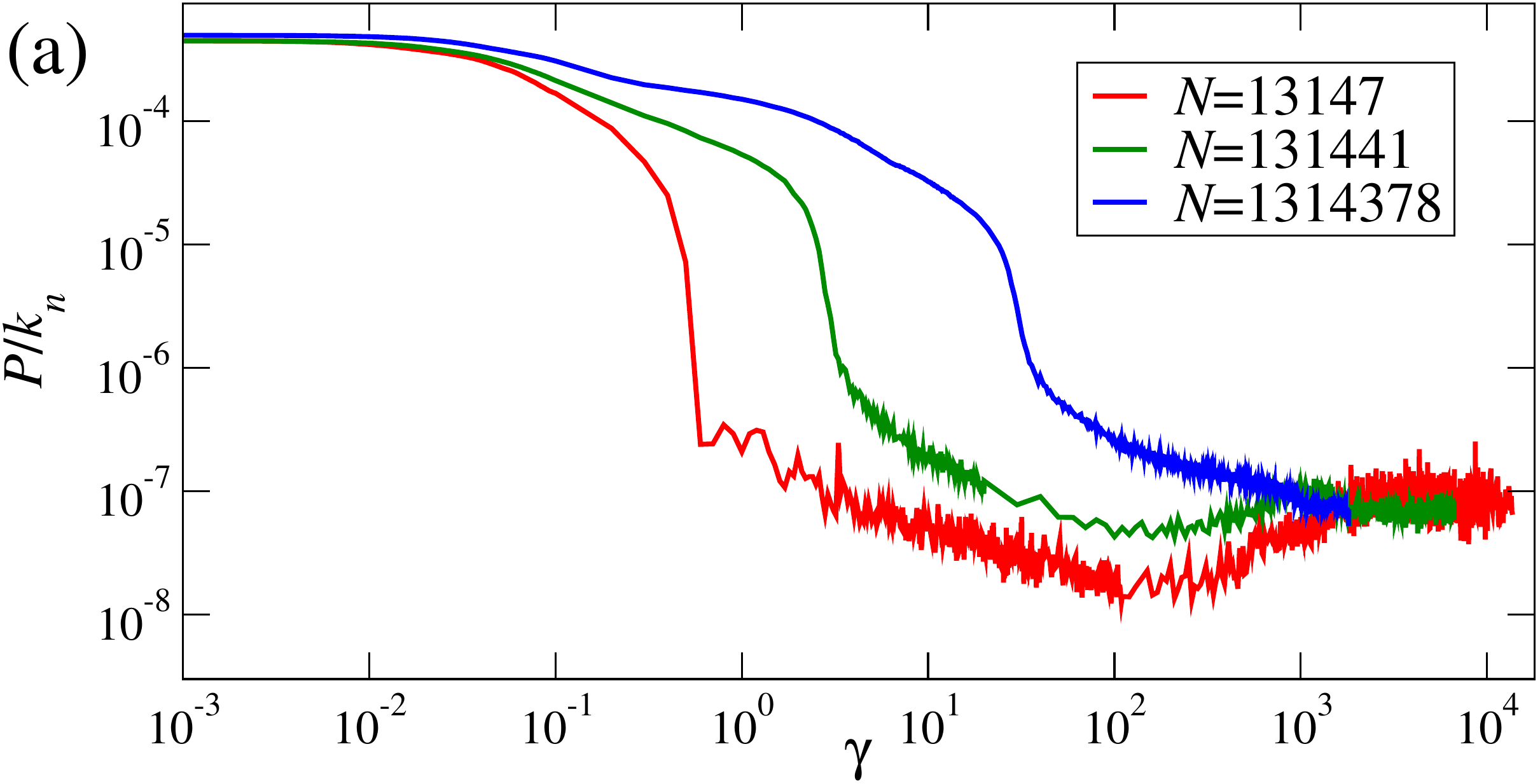}\\
        \includegraphics[width=0.4\textwidth]{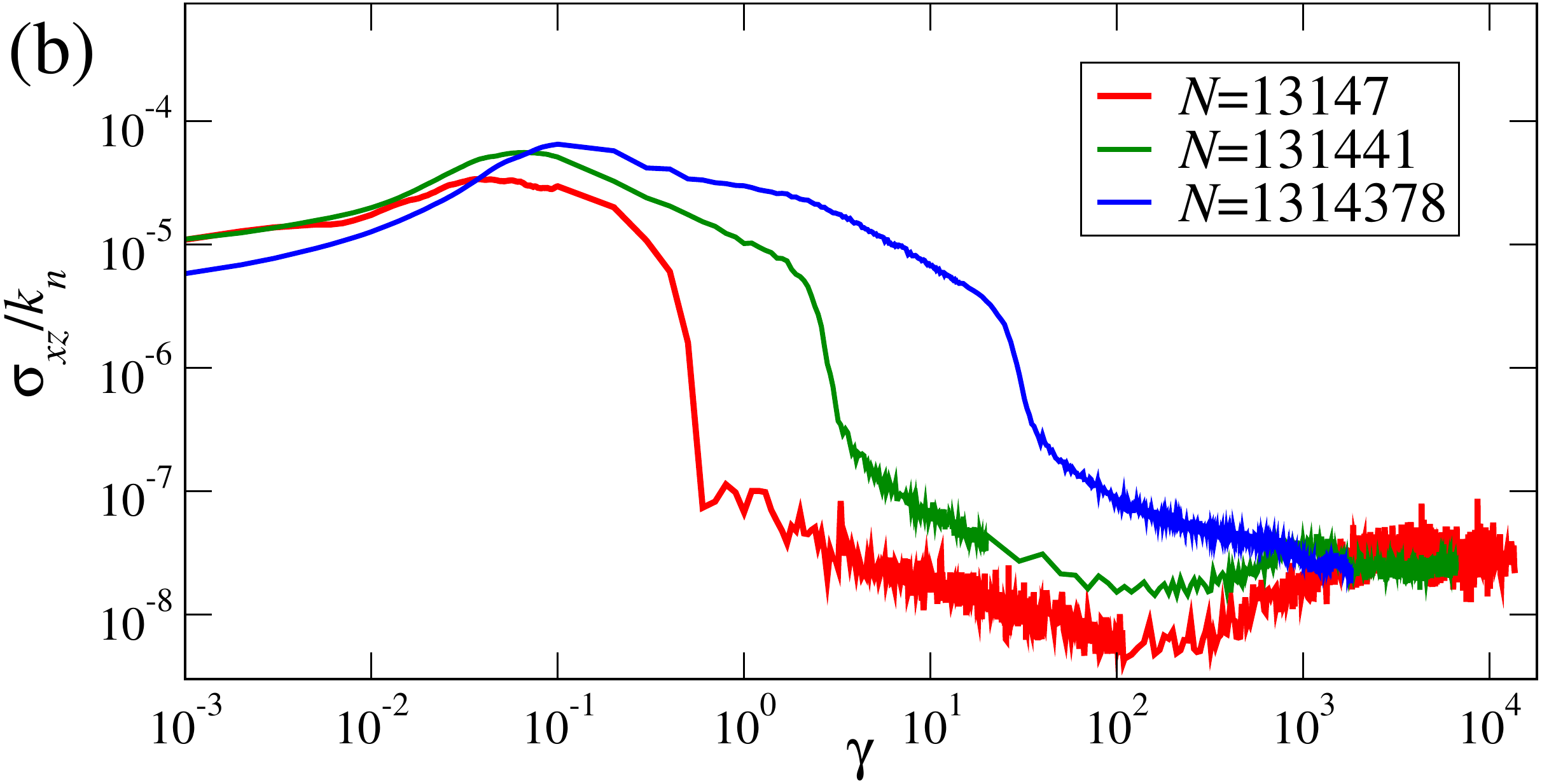}
        \caption{$P/k_n$ (a) and $\sigma_{\R{xz}}/k_n$ (b) over $\gamma$ at $\dot\gamma t_\R{col}=2.4\times 10^{-5}$ for several system sizes $N$.  
        }
        \label{Fig:L}
    \end{center}
\end{figure}
In this section, we conduct a finite-size analysis to understand how $P$, shear banding and mixing scenario are affected by the finite-size. 
The finite-size analysis involves three granular systems, whereby the smallest system contains more than ten thousand particles and the largest system contains more than one million particles. For the volume fraction of 0.6, the respective dimension of the three simulation systems are $L=$ 54.241, 114.94 and 247.26. We also studied another granular system with $L=28.728$, i.e., 34$\%$ greater than the largest particle diameter, containing 1318 particles. However, the results of such a small system is distinctly different from the rest and as such it is discarded.

We find that the pressure of our initial unmixed system, resting at mechanical equilibrium, is the same for all of the systems (Fig.~\ref{Fig:L}(a)). Here, we find qualitatively a similar response of the pressure against strain behavior irrespective of the system sizes. A significant pressure drop, which marks the mobility of the system, is prevalent in all of the three systems. The occurrence of pressure drop becomes significantly delayed with increasing system size. The system later gradually reaches steady states. We find that the steady state values of $P$ have negligible dependence on the system size. A similar significant drop and negligible dependence of the steady state values on the system size for $\sigma_{\R xz}$ are also shown in Fig.~\ref{Fig:L}(b).

\begin{figure}
    \begin{center}
        \includegraphics[width=0.4\textwidth]{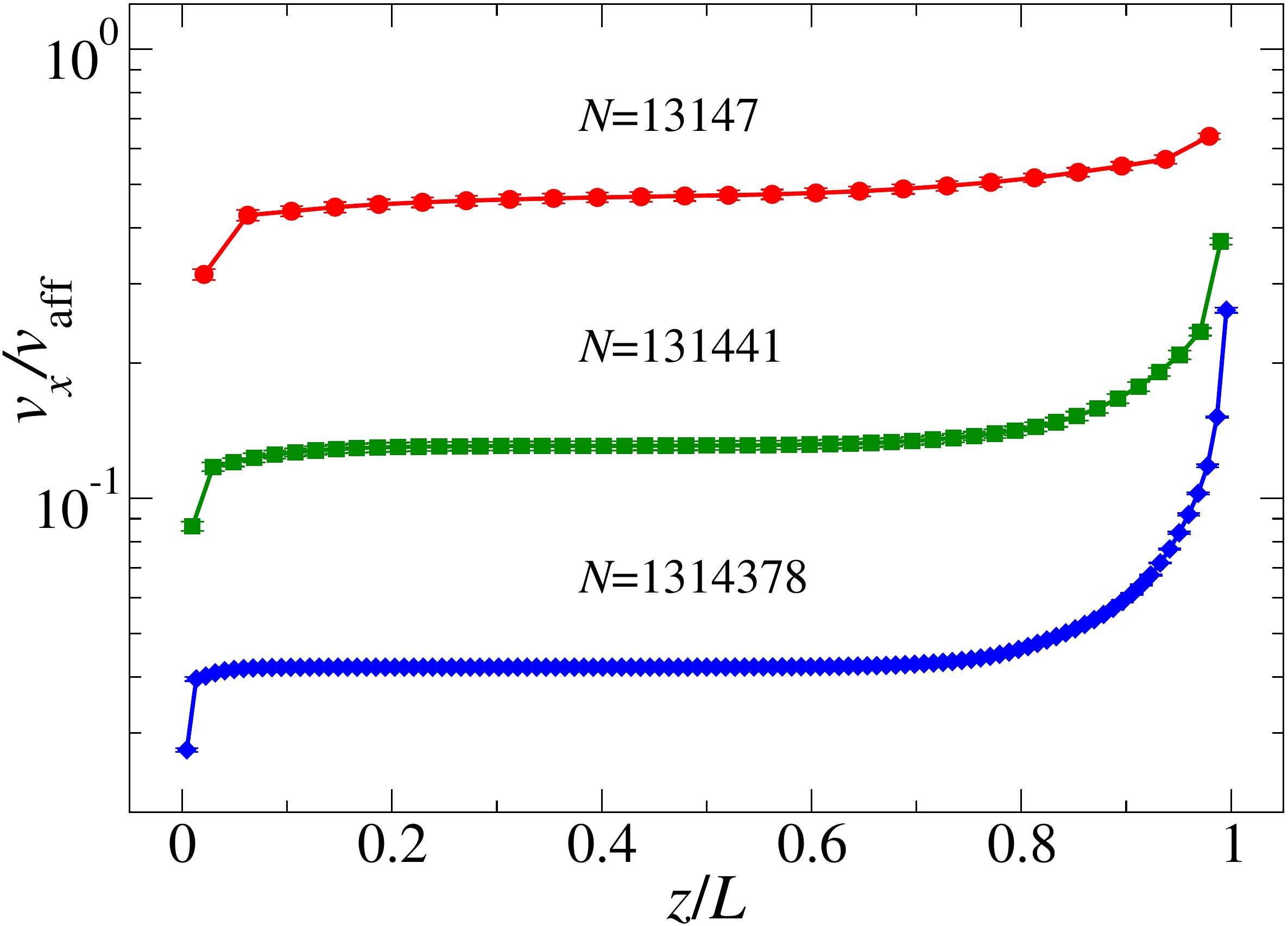}
        \caption{$v_x$ rescaled by $v_{\R{aff}}$ is plotted against $z/L$ at steady states for $\dot\gamma t_\R{col}=2.4\times 10^{-5}$ for several system sizes $N$. Error bars are smaller than the symbols.  
        }
        \label{Fig:zvxL}
    \end{center}
\end{figure}
Two shear banding regimes with a flat middle regime in $v_x$ profile are also prevalent for all of the system sizes over all values of $\gamma$. $v_x/v_{\R{aff}}$ with $z/L$ at steady states, is shown in Fig.~\ref{Fig:zvxL}. 
The steady state properties are averaged over the configurations saved in the last 300 strain intervals.
Two marked system size effects are observed: (i) the top regime becomes broader and (ii) the magnitude of $v_x/v_{\R{aff}}$ drops with increasing $N$. 

\begin{figure}
    \begin{center}
        \includegraphics[width=0.4\textwidth]{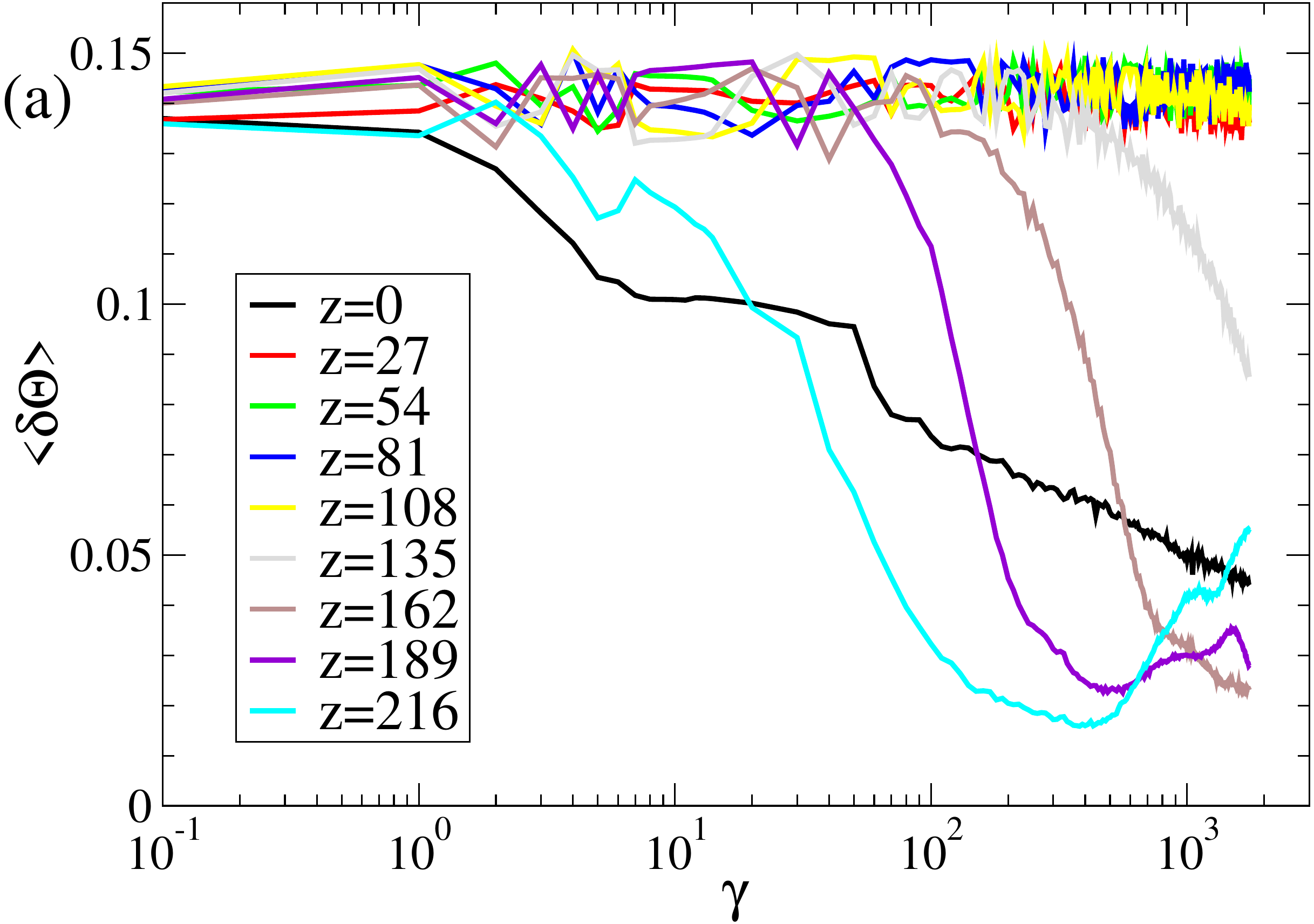}\\
        \includegraphics[width=0.4\textwidth]{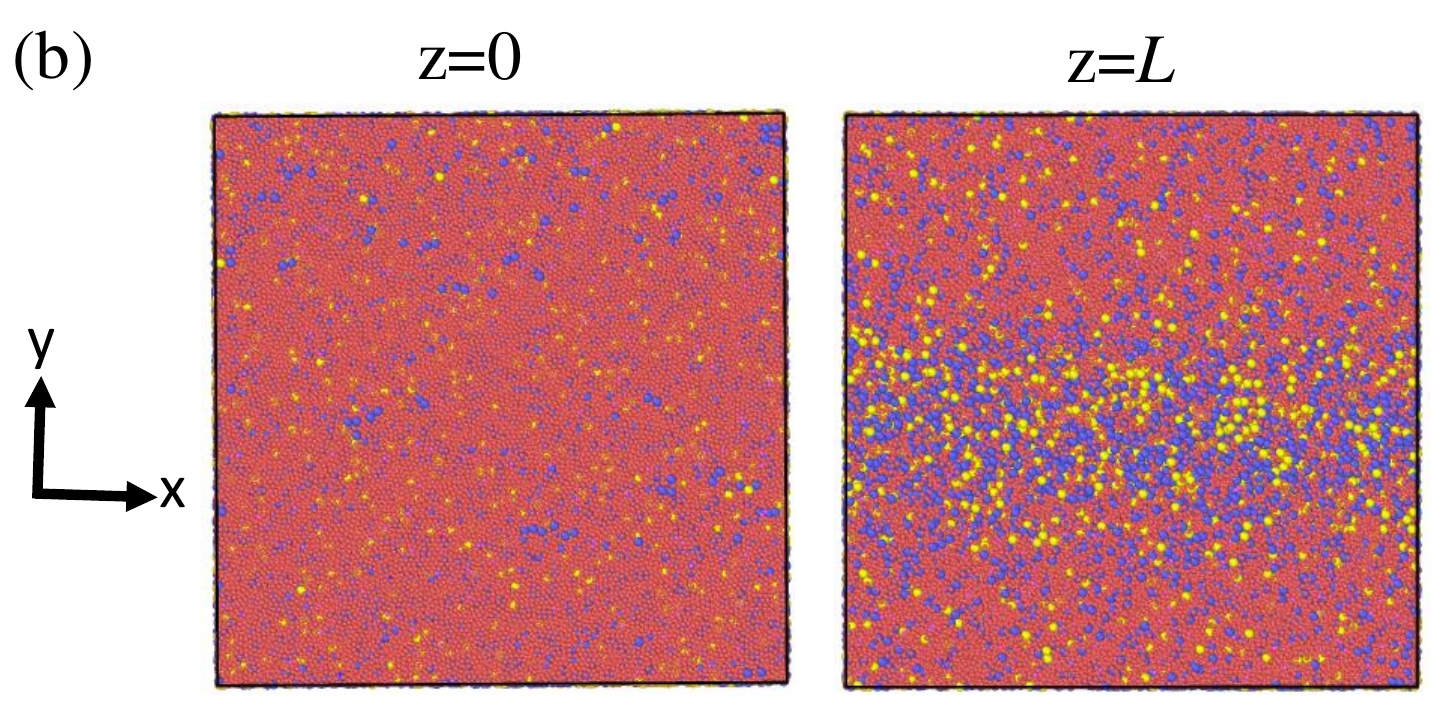}
        \caption{$\langle\delta\Theta\rangle$ versus $\gamma$ (a). The bottom plane (left) and the top (right) plane of the final configuration (b). Here $\dot\gamma t_\R{col}=2.4\times 10^{-5}$ and $N=1314378$.
        }
        \label{Fig:dohL}
    \end{center}
\end{figure} 
To understand the finite-size effect on mixing we plot $\langle{\delta\Theta}\rangle$ for the whole range of z-planes of the largest system as shown in Fig.~\ref{Fig:dohL}(a). 
$\langle{\delta\Theta}\rangle$ at the top and the bottom regimes drop fast compared to the bulk. 
In the bulk, the occurrence of drops is systematically delayed with decreasing z.
Note that the four regimes, just above $z=0$, do not show any decay. In order to achieve the decay we need to run longer simulations which are extremely time consuming. 
Nevertheless, a qualitatively similar mixing behavior is recovered as found in smaller systems.
Interestingly for our largest system, we observe a partial segregation. Granules of diameters $\sim 3-4$ accumulate more at the top wall compared to the other parts of the system (Fig.~\ref{Fig:dohL}(b)).

\subsection{Effects of strain rates}
\begin{figure}
    \begin{center}
        \includegraphics[width=0.4\textwidth]{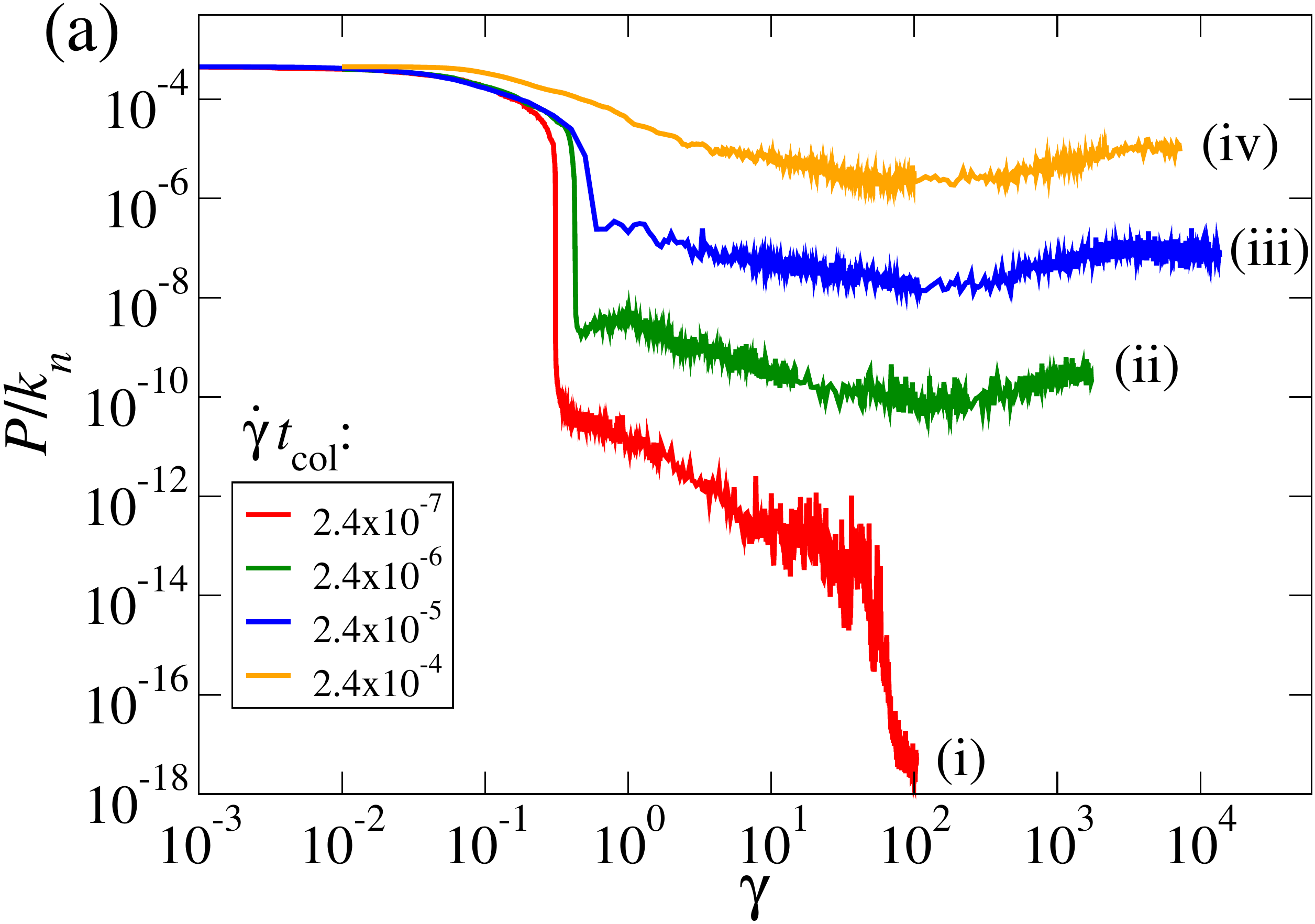}\\
        \includegraphics[width=0.4\textwidth]{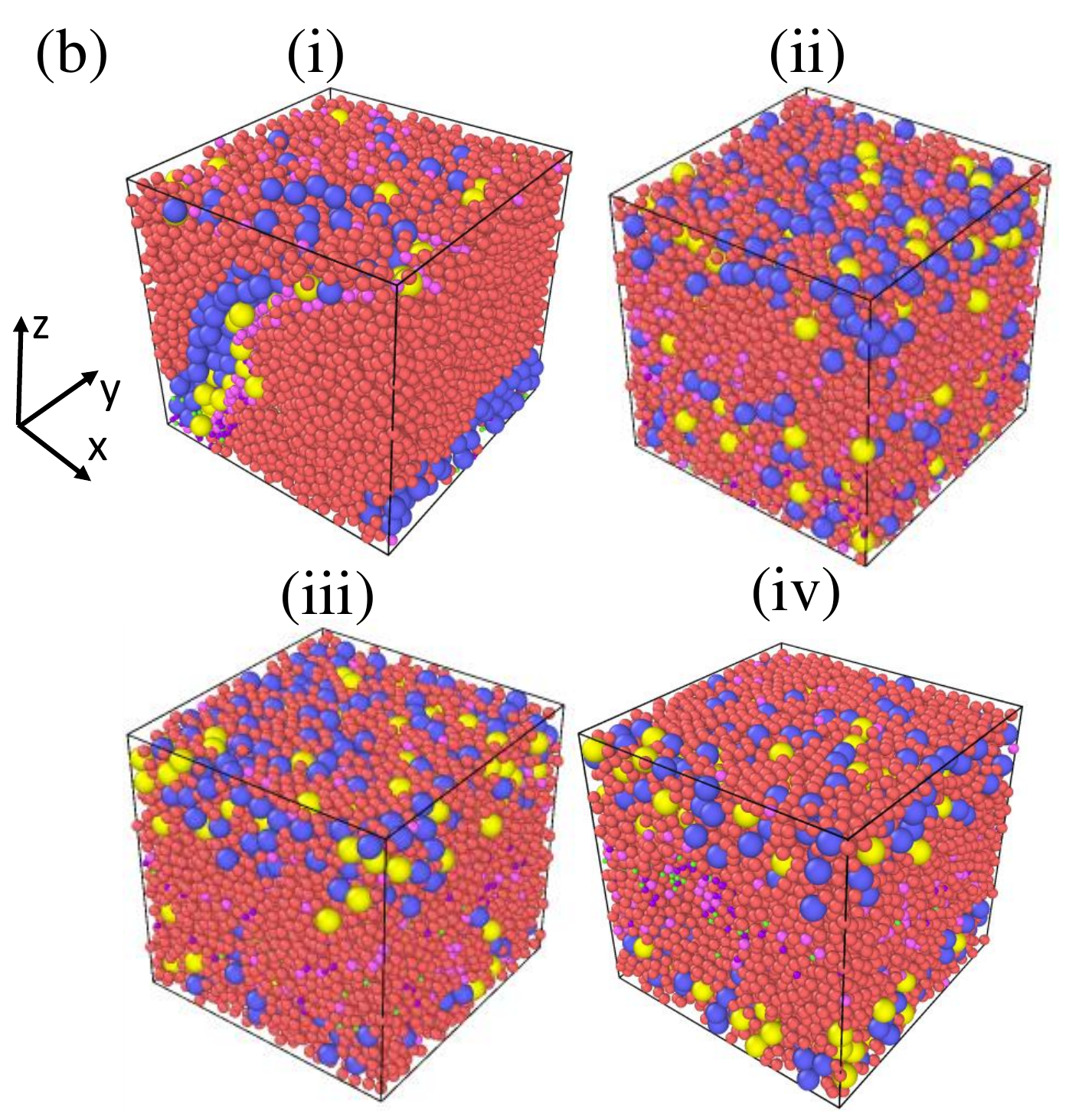}\\
        \includegraphics[width=0.4\textwidth]{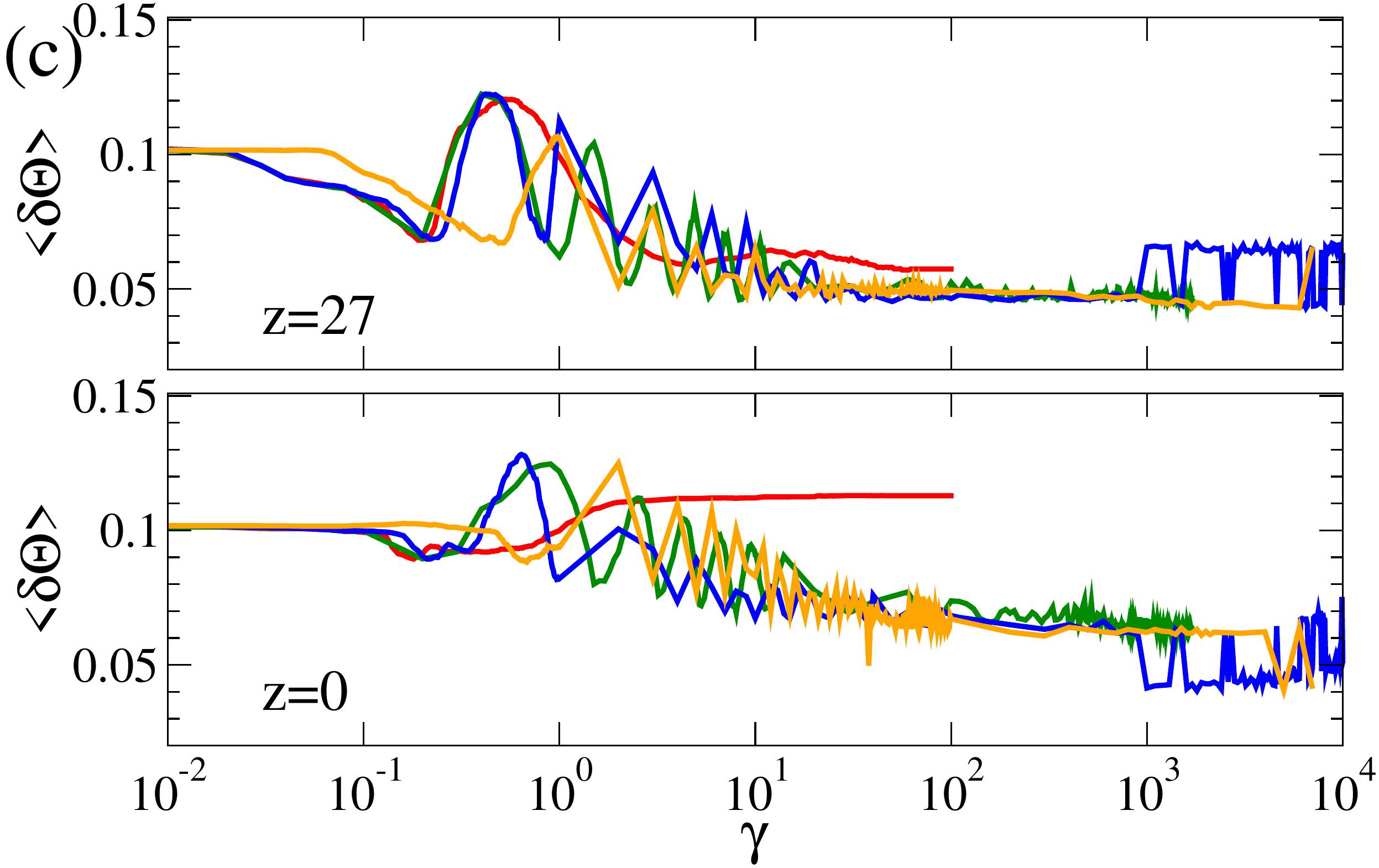}
        \caption{$P/k_n$ versus $\gamma$ for several strain rates and $N=13147$ (a), respective final configurations as shown in roman numerals (b). $\langle\delta\Theta\rangle$ versus $\gamma$ at z=0 (c, bottom panel) and z=27 (c, top panel) for the same strain rates. Here the box length $L=54.241$.
        }
        \label{Fig:gdot}
    \end{center}
\end{figure}
In this section, we study the effect of strain rate $\dot\gamma$ on mixing, especially, the correlation between $P$ and $\langle\delta\Theta\rangle$.
A strong dependence of $P$ on $\dot\gamma$ is evident in Fig.~\ref{Fig:gdot}(a). Note that $\dot\gamma$ is altered in this study by changing the value of $v_\R{aff}$. We find that the sharp drop in $P$, also observable in the previous section, decreases with increasing $\dot\gamma$. At the highest $\dot\gamma$ the drop becomes insignificant, which is in stark contrast at the lowest $\dot\gamma$ where the drop is prolonged. Interestingly, at these small values of the strain rate, the system reaches the zero pressure state over straining where no particles are in contact with the others including the walls.       

\begin{figure}
    \begin{center}
        \includegraphics[width=0.4\textwidth]{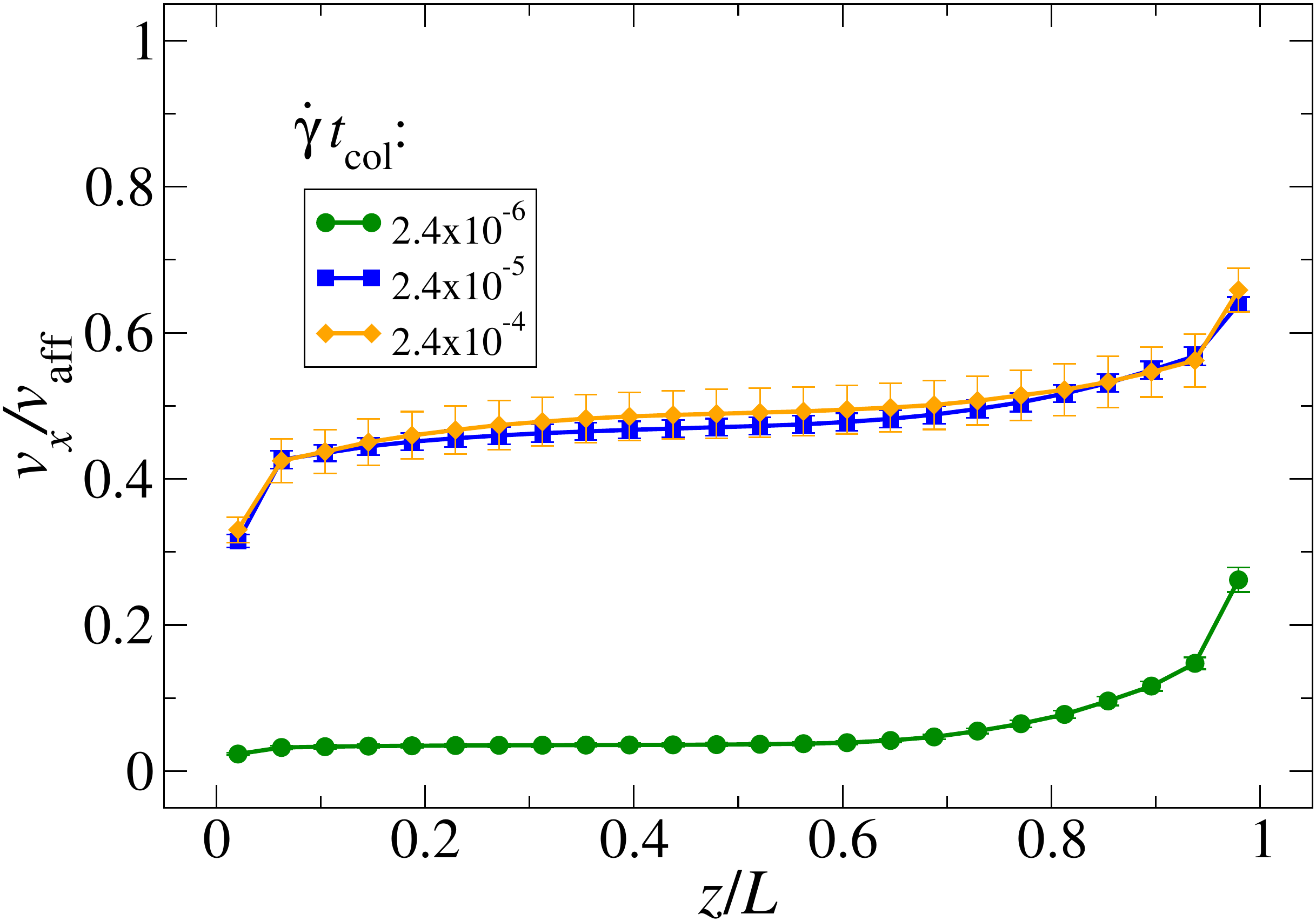}
        \caption{$v_x/v_{\R{aff}}$ versus $z/L$ at steady states for several strain rates and $N=13147$.
        }
        \label{Fig:zvxgdot}
    \end{center}
\end{figure}

Zero pressure results in no flows in the system, and thus, a homogeneous mixed phase cannot be achieved. A nearly unmixed final configuration for the lowest $\dot\gamma$ is shown in Fig.~\ref{Fig:gdot}(b). The granular particles reach the mixed phases for other high strain rates near the steady states. $\langle\delta\Theta\rangle$ for two regimes $z=0$ and $z=27$, where one regime lies on the bottom wall, whilst the other lies close to the top wall, are shown in Fig.\ref{Fig:gdot}(c). Initially, $\langle\delta\Theta\rangle$ remains constant, later it displays an oscillatory pattern. The oscillation is originated from shear deformation, implying a mass of particles periodically moving about in the same position in the box. The oscillation dies down with increasing $\gamma$ as the initial mass of the particles further mix with each other, creating a strain-independent mixing phase. As expected, for the lowest strain rate, we do not observe such mixing dynamics. At the mixed phases, which also coincide with the steady states, we observe permanent shear bands for all of the strain rates (Fig.~\ref{Fig:zvxgdot}). Further upon increasing strain rate, $v_x/v_{\R{aff}}$ collapses to a master curve pointing to the microscopic dynamics being identical at high strain rates. 

\subsection{Steady state properties}

\begin{figure}
    \begin{center}
        \includegraphics[width=0.45\textwidth]{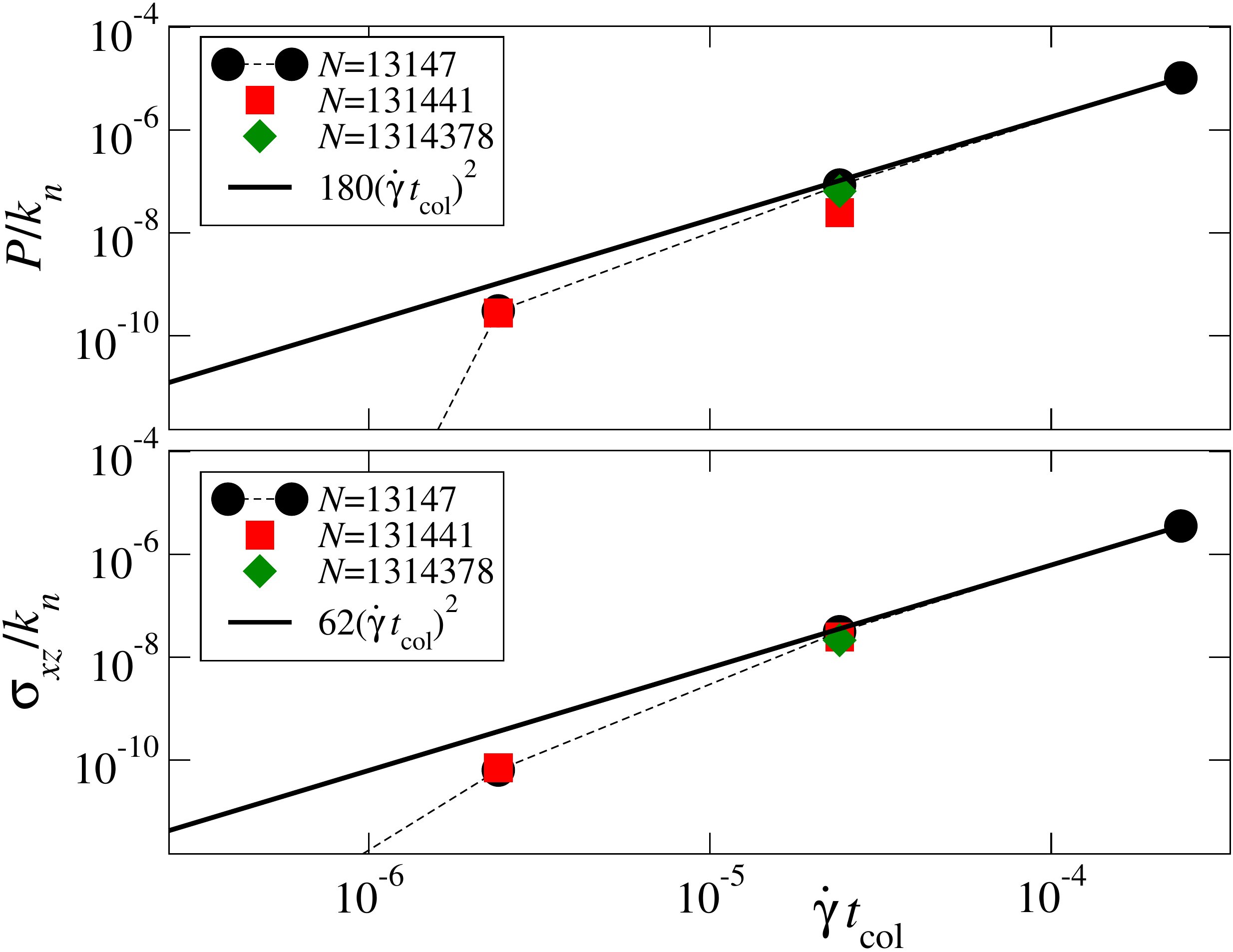}
        \caption{Steady state values of $P/k_n$ and $\sigma_{xz}/k_n$ as a function of $\dot\gamma t_{\R{col}}$.
        }
        \label{Fig:Pvsgdot}
    \end{center}
\end{figure}
In this section, we investigate the effect of mixing on the steady state properties. The steady state values of $P$ and $\sigma_{xz}$ are shown in Fig.~\ref{Fig:Pvsgdot}. Both the quantities asymptotically follow the scaling of $P \sim 180 t_{\R{col}}^2 \dot\gamma^2$ and $\sigma_{xz} \sim 62 t_{\R{col}}^2 \dot\gamma^2$. The quadratic scaling for stresses was first observed by Bagnold in dense granular flows~\cite{Bagnold1954}, which he rationalized using a kinetic theory where both, the frequency of binary collision and the momentum change per collision are assumed to be proportional to the strain rate. Later from his studies, Campbell concluded with the identification of three distinguished regimes in granular flows, namely, quasistatic, purely inertial and elastic inertial~\cite{Campbell2002}. Cialvo et al~\cite{ChialvoSS2012} then showed that stresses below the critical volume fraction, i.e., the jamming point $\phi_c$, belongs to a purely inertial regime where kinetic dynamics control the granular flow. 
In accordance with Cialvo et al, we also find that the current shear-induced mixing process results in a steady granular flow corresponding to a purely inertial regime. 
To further confirm the flow regime of the steady states,  
we estimate the relaxation time $\tau_{\R r}$ of the granular systems by letting one of our final configurations to relax in the absence of external shear. As a consequence, all the macroscopic variables decay to zero. 
Further, we fit pressure by $ P(t) \sim \exp(-t/\tau_{\R r})$, and find $t_{\R{col}}/\tau_{\R r}\approx 2\times 10^{-6}$.
Here, $t_{\R{col}}/\tau_{\R r}$ is served as a critical strain rate. 
When the strain rate is below the critical value, the flow is quasistatic, and it has no dependence on the strain rate values. 
However, when the strain rate is above the critical value the flow is purely inertial, which is consistent with the results in Fig.~\ref{Fig:Pvsgdot}. 

\section{Summary}
In summary, we found that the transition dynamics of mixing is sensitive to macroscopic dynamics. Starting from an unmixed phase, it was shown that the granular system attains a mixing phase typically at the similar strains at which the granular system reaches steady states. 

From the finite-size analysis we discovered a minimum system size, typically twice the size of the largest particle, above which the response of pressure and shear stress over strain and strain rates are qualitatively similar. The steady state values display little dependence on the system size, whilst the strain value, which marks the static to dynamic transition associated with a contact stress drop, significantly increases with the  system size.  

Mixing behavior is also qualitatively similar for all of the system sizes. We observed that the mixing dynamics is sensitive to shear band formation. Granules move fast around the mobile wall and become slow around the static wall that creates the three flow regimes at the top, bottom and bulk. The amount of shear required to mix granules in the bulk regime is significantly large as compared to mixing the granules in the two regimes close to  the walls, making the shear-induced mixing a slow process.

The steady state values of pressure and shear stress significantly depend on the strain rate, especially, we observed the Bagnold's scaling as found in purely inertial granular flow. A critical strain rate was found which marks the transition from the quasistatic to the purely inertial regime. Thus, any strain rates below the critical value, the system will have zero pressure, and as a result, it never reaches a mixed phase.   

Overall, our results suggest that elementary mechanisms underlying granular flows can be understood from a relatively small system, which might further lead one to construct a continuum model to quantify the mixing process at industrially relevant scales. 
Elementary mechanisms of mixing, especially the correlation between mixing and shear bands presented here should serve as a useful basis for future research. 
It would be interesting to study how non-Hertzian contact forces including cohesive forces and the particle shapes affect the mixing mechanisms as it is known in the literature that the formations of shear bands depend on the intensity of cohesive forces~\cite{Singh2014} as well  
as the particle shapes~\cite{Tian2020}. 
In addition, non-spherical particles, e.g., ellipsoids in rotating drums display different transverse mixing behaviors with the imposed rate in comparison with the spherical particles~\cite{HeGPZ2017}, which could also be a potential future research topic in the context of simple shear geometry.    

\section*{Acknowledgment}
Authors gratefully acknowledge the funding from RIE2020 AME IAF-PP grant with grant number A19C2a0019 and NSCC for allocating computation time and data storage. All the authors thank Lee Mun Wai, Tey Jayren, Thaddie Natalaray, Yap Fung Ling, Zhang Yong Wei, Liew Jun Xian for their valuable comments and discussion.   


\bibliographystyle{unsrt}


\end{document}